\documentclass[5p]{elsarticle}
\makeatletter
\def\ps@pprintTitle{%
 \let\@oddhead\@empty
 \let\@evenhead\@empty
 \def\@oddfoot{\centerline{\thepage}}%
 \let\@evenfoot\@oddfoot}
\makeatother

\usepackage[utf8]{inputenc}
\usepackage[english]{babel}
\usepackage[babel]{csquotes}
\usepackage{stmaryrd}
\usepackage{amssymb}
\usepackage{amsmath}
\usepackage{amsthm}
\usepackage{graphicx}
\newtheorem*{model}{Renormalised Brane World (RBW)}{\bf}{\it}
\newtheorem*{rem}{Remark}{\bf}{\it}

\newcommand{\li}[1]{\ensuremath{{^{(5)}}#1}}

\begin{document}

\begin{frontmatter}
\title{Standard Cosmology on the Anti-de Sitter boundary}

\author{Christian Henke} 
\ead{henke@math.tu-clausthal.de}

\address{University of Technology at Clausthal, Department of Mathematics,\\ Erzstrasse 1, D-38678 Clausthal-Zellerfeld, Germany}
%\institute{University of Technology at Clausthal, \email{henke@math.tu-clausthal.de},\\ Department of Mathematics,\\ Erzstrasse 1, D-38678 Clausthal-Zellerfeld, Germany}

\date{\today}

\begin{abstract}
The well-known brane world model of cosmology is an alternative to the standard model of cosmology, called $\Lambda$CDM (Lambda Cold Dark Matter) model. However, the quadratical energy density of the brane universe requires an undesirable fine tuning.
The paper develops a new Anti-de Sitter (AdS) brane world model without the quadratical energy density by using a second Einstein equation, which arises canonically during a holographic renormalisation.
On the AdS boundary it is demonstrated that this brane model equals an effective de Sitter $\Lambda$CDM model, whereas the brane model near the AdS boundary agrees to the $\Lambda$CDM model except the earliest times. 
Finally, it is shown that the coincidence problem is avoided and the cosmological constant problem, if not solved, is greatly reduced.  
%\PACS{
%       {95.30.Sf}{Relativity and gravitation} \and
%       {95.36.+x}{Dark energy} \and
%       {98.80.-k}{Cosmology} \and
%       {98.80.Jk}{Mathematical and relativistic aspect of cosmology} 
%}
\end{abstract}

%\maketitle
\end{frontmatter}

\section{Introduction}
One of the key open problems in physics is the scale difference of 121 orders of magnitude between the vacuum energy density of quantum field theory and the observed energy density of an empty universe (cosmological constant problem). Therefore, there are two ways to address the problem. On the one hand with the help of quantum field theory (see \cite{Martin_Everything_you_always_2012}) and, on the other hand, with a model of cosmology (see \cite{Henke_QuantumVacuumEnergy_2018,Henke_TheCosmologicalNature_2018,Henke_Cosmological_Inflation_2018}).
Taking both aspects into account is difficult, since gravity can probably only connected to quantum field theories in a space-time with negative curvature (Anti-de Sitter space-time), while the observable universe requires a space-time with positive curvature (de Sitter space-time).

Appropriate alternatives are brane world models which are generally based on negative cosmological constants (see \cite{Brax.vandeBruck.ea_Brane_world_cosmology_2004}). Unfortunately, they contain a quadratical energy density that can only be eliminated by an undesirable fine tuning.

The novelty of this work is to present an Anti-de Sitter (AdS) brane model that avoids the quadratical energy density and the related problems by an application of the holographic renormalisation. This renormalisation process reduces the boundary stress tensor and therefore the density to a finite energy scale (cf. \cite{Balasubramanian.Kraus_StressTensor_1999,Natsuume_AdSCFT_Duality_2015}).
Moreover, it is demonstrated that the brane model under consideration 
can describe cosmological observations at least as well as the $\Lambda$CDM model,
avoids the coincidence problem and greatly reduces the cosmological constant problem.

Due to the last points, one can conclude that the proposed brane model has a greater evidence than the $\Lambda$CDM model and therefore one can conclude that we live in an AdS space-time.
 
The remainder of the paper is organised as follows. In section 2 we review the well known brane cosmology and mention the restrictions of this approach. The next section reviews the holographic renormalisation and considers the associated action and variational principle only from the gravity point of view. In section 4 the Renormalised Brane World (RBW) equation is derived.
Then section 5 discusses the cosmology on the AdS boundary (conformal boundary).
In section 6, the ability of the RBW model to describe cosmological observations is investigated by analysing the deviations from the standard model of cosmology.
Finally, section 7 is devoted to concluding remarks. 

\section{Cosmological evolution on the brane}
\label{sec:cos_brane_1}
Let $G$ be the metric of the bulk space-time $\mathcal{M}=\mathcal{M}_d \times (y_{\text{br}},\infty)$ with a boundary $\partial \mathcal{M}=\mathcal{M}_d \times \{y_{\text{br}}\}.$ 
Using Gaussian normal coordinates any asymptotically AdS metric can be written in the form 
\begin{equation*}
ds^2=G_{AB}\, dx^A dx^B=\gamma_{\mu\nu}(x,y)\, dx^\mu dx^\nu+dy^2,
\end{equation*}
where the capital Latin letters are used for bulk indices and the Greek alphabet for $d$ space-time indices.
The bulk space-time can be seen as a family of foliated timelike hypersurfaces which are labeled by their coordinate $y.$
Especially, the cosmological evolution on the brane can be described by the metric of Robertson and Walker (RW) 
\begin{equation}
ds^2=-n^2(t,y) dt^2+a^2(t,y) d\Omega_3^2 + dy^2,\quad n(t,0)=1,
\label{eq:RW_metric}
\end{equation}
where the line element is given by
\begin{equation*}
d\Omega_3^2=\frac{dr^2}{1-kr^2}+r^2 \left( d\theta^2+\sin^2(\theta) d\phi^2 \right),
\end{equation*}
and $k$ is referred as the intrinsic curvature parameter.
Now, Einstein's field equation leads to the modified Friedmann equation (cf. \cite{Martens.Koyama_Braneworld_Gravity_2010,Binetruy.Deffayet.ea_Brane_cosmological_evolution_2000,Brax.vandeBruck.ea_Brane_world_cosmology_2004}) 
\begin{equation}
-\frac{1}{3}\Lambda_5 +\frac{k}{a^2} +\left(\frac{\dot{a}}{a n}\right)^2 
=\left( \frac{\kappa_5^2 \tilde{\rho}}{6} \right)^2 +\frac{\mu}{a^4}-\frac{\Lambda_5}{6},\quad \kappa_5^2=8 \pi G_5,
\label{eq:mod_friedmann}
\end{equation}
where $\Lambda_5<0$ is the cosmological constant, $G_5$ is the 5-dimensional gravitational coupling constant and $\mu$ is a free parameter of the dark radiation term $\frac{\mu}{a^4}.$
Usually, it is assumed that the hypersurface at $y_{\text{br}}=0$ forms our universe. 

Moreover, $\tilde{\rho}$ denotes the density of the energy-momen\-tum tensor $\tilde{T}_{\mu\nu}$ 
\begin{equation}
\tilde{T}_{\mu\nu}=- \frac{2}{\kappa_5^2}\left(K_{\mu\nu}-K \gamma_{\mu\nu}\right), \quad K=K_{\mu\nu} \gamma^{\mu\nu},
\label{eq:T_matt}
\end{equation}
which is defined by Israel's junction condition (see \cite{Israel_Singular_hypersurfaces_thin_shells_1966}). 
%Moreover, $-\tilde{T}_{\mu\nu}/2$ is also known as the quasi-local stress-energy tensor of Brown and York (see \cite{Brown.York_Quasilocal_energy_1993}).
The extrinsic curvature is given by $K_{\mu\nu}=\frac{1}{2} \mathcal{L}_n \gamma_{\mu\nu},$ where $\mathcal{L}_n$ is the Lie derivative w.r.t. the inward-pointing unit normal vector $n_\mu.$ 
The left hand side of $(\ref{eq:mod_friedmann})$ consists of the terms that occur in the usual Friedmann equation, whereas the right hand side includes the terms that are caused by the brane context. Especially the quadratical energy density is initially incompatible to the standard model of cosmology and requires a decomposition of the form 
\begin{equation*}
\tilde{\rho}=\rho_{\Lambda}+\rho,\;
%\rho \ll |\rho_{\Lambda}|,\;
G_4=\frac{\kappa_5^4}{48\pi} \rho_{\Lambda},\;
\Lambda_4=\frac{\kappa_5^4}{12} \rho_{\Lambda}^2 +\frac{\Lambda_5}{2}.
\end{equation*}
Hence, it follows that 
\begin{equation*}
\left( \frac{\kappa_5^2 \tilde{\rho}}{6} \right)^2 +\frac{\Lambda_5}{6}
=\frac{8\pi G_4}{3}\rho+ \left(\frac{\kappa_5^2}{6}\right)^2  \rho^2
+\frac{\Lambda_4}{3},
\label{}
\end{equation*}
where $\rho_\Lambda$ is the tension of the brane.
Therefore, the standard model of cosmology is obtained for $\mu=0$ and if the low energy case $\rho \ll 96 \pi G_4/\kappa_5^4$ is considered. 
%For a brief overview about the constraints and problems of this approach see \cite{Brax.vandeBruck.ea_Brane_world_cosmology_2004} and \cite{Shiromizu.Maeda.ea_Einstein_3brane_world_2000}. 
However, in order to keep the observational quality of the $\Lambda$CDM model, the high energy range $\rho \gg 96 \pi G_4/\kappa_5^4$ must take place before nucleosynthesis. For this purpose it is necessary to use the fine tuning condition $\Lambda_4=0$ of the Randall-Sundrum model and require that $\rho$ includes the matter, the radiation and the dark energy densities of the universe (cf. \cite{Brax.vandeBruck.ea_Brane_world_cosmology_2004,Cline.Grojean.ea_Cosmological_Expansion_Presence_1999,Abdusattari.Iminniyaz_Abundance_Asymmetric_Dark_2016}).
In that case and using the settings $k=\mu=0,$ the Hubble expansion rate $H=\dot{a}/a$ yields
\begin{equation*}
H^2=H_{\Lambda CDM}^2 \left( 1+\frac{\kappa_5^4}{96 \pi G_4} \rho \right),
\label{}
\end{equation*}
where $H_{\Lambda CDM}$ denotes the corresponding Hubble rate of the $\Lambda$CDM model.
According to \cite{Verde.Treu.ea_Tensions_between_early_2019}, the increased Hubble rate and the fact that the deviations from the $\Lambda$CDM model concerns only the earliest epochs prepares the brane cosmology as a candidate to overcome the tensions between the early and late universe. This seems to be confirmed by the Planck Data from 2015, since the best fit values of the Hubble constant do not generate a tension between the early and late universe 
(see \cite{Garcia.Magana.ea_Probing_dark_energy_2018}).

One way to derive the modified Friedmann equation $(\ref{eq:mod_friedmann})$ is to start with five dimensional Schwarzschild-AdS coordinates and identify the fifth dimension with the scale factor $a(t).$ In other words, the brane is allowed to move in the bulk along the fifth dimension (cf. \cite{Martens.Koyama_Braneworld_Gravity_2010,Bilic_RandallSundrum_versus_holographic_2016}).
Besides the fine tuning, a second problem of equation 
$(\ref{eq:mod_friedmann})$ 
becomes visible at this point. 
Typically $\tilde{T}_{\mu\nu}$ and therefore $\tilde{\rho}$ diverges if the brane is moved to the conformal boundary (cf. \cite{Balasubramanian.Kraus_StressTensor_1999,de_Haro.Skenderis.ea_Holographic_Reconstruction_2001}). 

Further research of the brane world model $(\ref{eq:mod_friedmann})$ was done in the framework of the AdS/CFT correspondence, where a conformal field theory is defined on the AdS boundary,
 (cf. \cite{Apostolopoulos.Siopsis.ea_Cosmology_AdS_Schwarzschild_2009,Bilic_RandallSundrum_versus_holographic_2016}). However, these studies derive a common holographic Friedmann equation which is still based on the Randall-Sundrum action and can recover the standard cosmology only in an unnatural or fine tuned setting.

%In the context of the AdS/CFT correspondence, the divergences may be removed by 
%the holographic renormalisation (cf. \cite{Balasubramanian.Kraus_StressTensor_1999,de_Haro.Skenderis.ea_Holographic_Reconstruction_2001,Natsuume_AdSCFT_Duality_2015}) which is realised by adding special counterterms to the boundary action term. 
% which generates an increased Hubble rate in the earliest epochs and keep the observational quality of the $\Lambda$CDM model in the further course of the universe.

\section{Holographic renormalisation}
Throughout this paper, the sign convention are defined by
the Riemann
\begin{equation*}
R_{\mu\nu\alpha}^{\phantom{xxx}\beta}=-\left( \partial_\mu \Gamma_{\nu\alpha}^{\phantom{xx}\beta} 
+\Gamma_{\mu\gamma}^{\phantom{xx}\beta} \Gamma_{\nu\alpha}^{\phantom{xx}\gamma}-\mu \leftrightarrow \nu\right),
\end{equation*}
and the Ricci tensor $R_{\mu\nu}=R_{\mu\alpha\nu}^{\phantom{xxx}\alpha}, R=R_\mu^\mu.$
 
Since $\tilde{T}_{\mu\nu}$ diverges as the hypersurface approaches the conformal boundary, the energy-momentum tensor is renormalised by adding finite terms to $\tilde{T}_{\mu\nu}.$This can be done without disturbing the bulk equations by adding boundary terms to the action. 
In particular for $d=4$ 
this renormalised action (cf.  \cite{de_Haro.Skenderis.ea_Holographic_Reconstruction_2001}) can be written as
\begin{equation}
S=S_{\text{gr}}+S_{\text{ct}}
\label{eq:S_Haro}
\end{equation}
where
\begin{equation}
\begin{split}
S_{\text{gr}}&=\frac{1}{2\kappa_5^2} \int_{\mathcal{M}} \left( \li{R}-2\Lambda_5 \right) \sqrt{-G}\,d^5 x\\
&-\frac{1}{\kappa_5^2} \int_{\partial \mathcal{M}} K\sqrt{-\gamma}\, d^4 x, 
%+\int_{\mathcal{M}} L_{\text{bulk}} \sqrt{-G}\, d^5 x \\
%+S_m,\\
%S_m&=\int_{\partial \mathcal{M}} L_{\text{matter}} \sqrt{-\gamma}\, d^4 x. \\
\end{split}
\label{eq:einstein_hilbert}
\end{equation}
is the usual Einstein-Hilbert action including the Gibbons-Hawking boundary term and 
$S_{\text{ct}}$ cancels the divergent terms in $S_{\text{gr}}.$
$S_{\text{ct}}$ can be determined by an analysis of $S_{\text{gr}}$ on the conformal boundary  
\begin{equation*}
\begin{split}
S_{\text{gr}}&=\frac{l^3}{2 \kappa_5^2} \int_{\mathcal{M}} \left( 6 r^{-4} +a_{(4)}\log(r^2)+f(r) \right) \sqrt{-g_{(0)}}\,d^4 x.
\end{split}
%\label{eq:einstein_hilbert2}
\end{equation*}
Here, $f(r)=\mathcal{O}(r^0)$ and the coefficient $a_{(4)}$ is given by
\begin{equation*}
a_{(4)}=\frac{1}{2} \text{tr}\left( g_{(0)}^{-1}g_{(2)} \right)^2 -\frac{1}{2} \text{tr}\left( \left[ g_{(0)}^{-1} g_{(2)} \right]^2 \right),
\end{equation*}
where $g_{(0)}$ and $g_{(2)}$ come from the expansion 
\begin{equation}
\begin{split}
g(x,r)&=g_{(0)}(x)+r^2 g_{(2)}+r^4 g_{(4)}(x)\\
&+r^4 \log(r^2)h_{(4)}(x)+ \mathcal{O}(r^{6})
\end{split}
\label{eq:FG}
\end{equation}
of the Fefferman-Graham coordinates
\begin{equation}
ds^2=\frac{l^2}{r^2}\left( g_{\mu\nu}(x,r)\, dx^\mu dx^\nu +dr^2\right),\quad r=l e^{y/l}.
\label{eq:coord_Feff_Gra}
\end{equation}
Rewriting $S_{\text{gr}}$ in terms of the induced metric $\gamma$ gives
\begin{equation}
\begin{split}
S_{\text{gr}}&=\frac{1}{2l \kappa_5^2} \int_{\mathcal{M}} \left(
6 +l^2 \frac{R}{2}+r^4 \log\left( r^2 \right) a_{(4)} \right) \sqrt{-\gamma}\,d^4 x\\
&+\frac{1}{2l \kappa_5^2} \int_{\mathcal{M}} r ^4 \left(
\frac{1}{2}a_{(4)}-f(r)
 \right) \sqrt{-\gamma}\,d^4 x\\
&+\frac{1}{2l \kappa_5^2} \int_{\mathcal{M}} \left(
\mathcal{O}(R^3 l^6 \log(r^2))
 \right) \sqrt{-\gamma}\,d^4 x.
\end{split}
\label{eq:einstein_hilbert2}
\end{equation}
To regulate the theory, the action is evaluated at the cut-off $r_{\text{br}}=le^{y_{\text{br}}/l}$ 
which should be close to the conformal boundary at $r=0.$
Provided that 
\begin{equation}
|R l^2 \log(r_\text{br}^2)| \ll 1,
\label{eq:l_const}
\end{equation}
the contributions of the remainder term in equation $(\ref{eq:einstein_hilbert2})$ are small compared to the first term of this equation.
Therefore, the counterterm action of a minimal subtraction scheme is given by
\begin{equation}
\begin{split}
S_{\text{ct}}&=-\frac{l}{2 \kappa_5^2} \int_{\partial \mathcal{M}}\left(\frac{6}{l^2} +\frac{R}{2}
%+ l^2\log\left( l^2 e^{2y/l} \right) \frac{R^2}{24}-\frac{R_{\mu\nu}R^{\mu\nu}}{8}
 \right) \sqrt{-\gamma}\, d^4x\\
&-\frac{1}{2 l\kappa_5^2} \int_{\partial \mathcal{M}} a_{(4)} r_{\text{br}}^4\log\left(r_{\text{br}}^2 \right)\sqrt{-\gamma}\, d^4x.
\end{split}
\label{eq:Sct}
\end{equation}
As shown in \cite{de_Haro.Skenderis.ea_Holographic_Reconstruction_2001} the variation $\delta S$ is finite on the boundary and defines a valid variational principle
\begin{equation}
\delta S=-\frac{1}{2} \int_{\partial \mathcal{M}} T_{\mu\nu} \delta \gamma^{\mu\nu} \sqrt{-\gamma}\, d^4x
\label{eq:dS_Haro}
\end{equation}
for some finite energy-momentum tensor $T_{\mu\nu}.$
Usually, the AdS/CFT correspondence is used at this point and $T_{\mu\nu}$ can be interpreted as the expectation value of the stress-energy tensor of the dual CFT 
(cf. \cite{de_Haro.Skenderis.ea_Holographic_Reconstruction_2001}). Besides the correspondence, $T_{\mu\nu}$ has the original meaning ``as the quasi-local stress-energy tensor of Brown and York'' (cf. \cite{de_Haro.Skenderis.ea_Holographic_Reconstruction_2001}) and ``characterises the entire system including contribution from both the gravitational field and matter fields'' (see \cite{Brown.York_Quasilocal_energy_1993}).

In this paper, the variation of the matter action is identified with the renormalised stress energy tensor $T_{\mu\nu}.$
This can be done with the Lagrangian density $L_{\text{matter}}$ by the setting 
\begin{equation}
\begin{split}
T_{\mu\nu}&=-\frac{2}{\sqrt{-\gamma}} \frac{\delta S_m}{\delta \gamma^{\mu\nu}},\\
S_m&=\int_{\partial \mathcal{M}} L_{\text{matter}} \sqrt{-\gamma}\, d^4 x. \\
\end{split}
\label{eq:T_munu_def2}
\end{equation}
The variational principle $(\ref{eq:dS_Haro})$ can be written as
\begin{equation}
\begin{split}
&\frac{1}{2\kappa_5^2} \int_{\mathcal{M}} \li{\pi}_{AB}  \delta G^{AB} \sqrt{-G}\, d^5 x\\
&+\frac{1}{2\kappa_5^2} \int_{\partial \mathcal{M}}\pi_{\mu\nu} \delta \gamma^{\mu\nu} \sqrt{-\gamma}\, d^4 x=0.
\end{split}
\label{eq:variation}
\end{equation}
Here, the following notations have been adopted: 
\begin{equation}
\begin{split}
\li{\pi}_{AB}&= \li{R_{AB}}-\frac{1}{2}G_{AB} \li{R}+\Lambda_5 G_{AB}, \\
\pi_{\mu\nu}&= \frac{l}{2}\bigg( E_{\mu\nu}-\frac{6}{l^2} \gamma_{\mu\nu} -4 r_{\text{br}}^2\log\left(r_{\text{br}}^2\right) h_{(4)\mu\nu}\bigg)\\
&-\kappa_5^2 \bigg( \frac{\tilde{T}_{\mu\nu}[\gamma]}{2}+ T_{\mu\nu}[\gamma]\bigg),\\
E_{\mu\nu} &=R_{\mu\nu}[\gamma] -\frac{1}{2}R[\gamma] \gamma_{\mu\nu}.
\end{split}
\label{}
\end{equation}
If one applies Dirichlet boundary conditions $\delta \gamma^{\mu\nu}=0$ equation $(\ref{eq:variation})$ is satisfied by 
\begin{equation}
\li{\pi}_{AB}=0,\; \text{in } \mathcal{M}.
\label{eq:einstein5a}
\end{equation}
Taking the trace of $(\ref{eq:einstein5a}),$ the cosmological constant $\Lambda_5=-6/l^2$ follows from the negative curvature of the AdS space-time $\li{R}=-20/l^2.$

In this work, dynamical metrics $\gamma_{\mu\nu}$ are considered. Therefore, equation $(\ref{eq:variation})$ is fulfilled by $(\ref{eq:einstein5a})$ and
\begin{equation}
\pi_{\mu\nu}=0,\; \text{in } \partial\mathcal{M}. 
\label{eq:einstein5b}
\end{equation}

Notice that, both equations $(\ref{eq:einstein5a})$ and $(\ref{eq:einstein5b})$ depend on $\tilde{T}_{\mu\nu}.$ The aim of this work is to determine and then remove the tilde terms from both equations. Thus, the quadratic energy density $\tilde{\rho}^2$ of equation $(\ref{eq:mod_friedmann})$ also disappears. 

The equations $(\ref{eq:einstein5a})$ and $(\ref{eq:einstein5b})$ are equivalent to the equations presented in \cite{de_Haro.Skenderis.ea_Gravity_warped_2001}.  The equations in \cite{de_Haro.Skenderis.ea_Gravity_warped_2001} are characterised by the fact that no counterterms were added to the action and that the variation of the matter action is identified with the infinite stress energy tensor $\tilde{T}_{\mu\nu}.$ The decomposition of $\tilde{T}_{\mu\nu}$ in a renormalised part and a counterterm remainder ensures the equivalence with the approach under consideration, where the counterterm is added to the action and the variation of the matter action is identified with the renormalised stress energy tensor $T_{\mu\nu}.$

Now, it is supposed that the boundary metric $g_{(0)}$ has a vanishing Bach tensor. Then, using the property that the Bach tensor is proportional to $h_{(4)\mu\nu}$ (see \cite{Genolini.Cassani.ea_Holographic_renormalization_2017}), the following Einstein equation 
\begin{equation}
E_{\mu\nu}+\Lambda_5 \gamma_{\mu\nu}=
\kappa_4^2 \left(T_{\mu\nu}+\frac{1}{2} \tilde{T}_{\mu\nu}\right), \quad \kappa_4^2=\frac{2\kappa_5^2}{l}
\label{eq:en_ten_cft}
\end{equation}
is satisfied on the brane.
The aim of the next section is to remove the quadratic energy density $\tilde{\rho}^2$ of equation $(\ref{eq:mod_friedmann})$ by combining the equations $(\ref{eq:einstein5a})$ and $(\ref{eq:en_ten_cft})$ and derive a brane cosmology without fine tuning.

\section{Renormalised brane cosmology}
In this section, the holographic renormalisation is applied to the brane cosmology of section \ref{sec:cos_brane_1}. First, it has to be ensured that the line element $(\ref{eq:RW_metric})$ has a vanishing Bach tensor on the boundary. 
This can be demonstrated by the following arguments: $\gamma(x,0)$ and hence $g_{(0)}$ are standard RW metrics, which are conformally flat (cf. \cite{Tauber_Expanding_Universe_1967}) and therefore conformally Einstein metrics. According to \cite{Kozameh.Newman.ea_Conformal_Einstein_spaces_1985}, the latter property is sufficient for a vanishing Bach tensor.

Now the Friedmann equations $(\ref{eq:mod_friedmann})$ and
\begin{equation}
-\frac{1}{3} \Lambda_5 +\frac{k}{a^2} +\left(\frac{\dot{a}}{a n}\right)^2
= \frac{ \kappa_4^2}{3} \left( \rho+\frac{1}{2} \tilde{\rho} \right),
\label{eq:cft_einstein}
\end{equation}
which follow from the equations $(\ref{eq:einstein5a})$ and $(\ref{eq:en_ten_cft})$ in the case of the line element $(\ref{eq:RW_metric})$ are investigated.
Here, the energy density $\rho$ is related to the renormalised tensor $T_{\mu\nu}.$
Solving the quadratical equation for $\tilde{\rho}$ which follows from the identification of the right hand sides of $(\ref{eq:mod_friedmann})$ and $(\ref{eq:cft_einstein})$ and reinserting the solution implies
\begin{equation}
\left(\frac{\dot{a}}{a n}\right)^2 +\frac{k}{a^2}
=\frac{\kappa_4^2}{3} \rho + \epsilon \frac{2 \sqrt{\kappa_4^2 \rho/3-\mu/a^4}}{l}, \quad \epsilon=\pm 1.
\label{eq:mod_friedmann2}
\end{equation}
Notice, that the renormalisation mechanism has 
eliminated the quadratical density term in equation $(\ref{eq:mod_friedmann}).$ However, another density term has appeared under the square root. In the following, it is demonstrated how this term fits into the standard cosmology.
Using todays Hubble constant $H_0,$ todays scale factor $a_0=a(t_0,y_{\text{br}}),$ the setting $x=a/a_0$ and $\rho_{\text{crit}}=3H_0^2/\kappa_4^2,$ equation $(\ref{eq:mod_friedmann2})$ can be written as
\begin{equation}
\frac{\left( \frac{d}{d\tau}x \right)^2}{x^{2}}-\frac{\Omega_k}{ x^{2}}=\frac{\rho}{\rho_{\text{crit}}} +2\epsilon \sqrt{-\frac{\Omega_{\Lambda_5}}{2}} \sqrt{\frac{\rho}{\rho_{\text{crit}}} -\frac{\Omega_{\mu}}{x^{4}}}, 
\label{eq:mod_friedmann3}
\end{equation}
or as 
\begin{equation}
\frac{\left( \frac{d}{d\tau}x \right)^2}{x^{2}}-\frac{\Omega_k}{ x^{2}}
= \frac{\Omega_{\Lambda_5}}{2}+\frac{\Omega_\mu}{x^4}
+\left( \sqrt{\frac{\rho}{\rho_{\text{crit}}} -\frac{\Omega_{\mu}}{x^{4}}} +\epsilon \sqrt{-\frac{\Omega_{\Lambda_5}}{2}} \right)^2, 
\label{eq:mod_friedmann4}
\end{equation}
where the additional variables are denoted by $\Omega_k=-k/(H_0^2a_0^2),$ $\Omega_{\Lambda_5}=\Lambda_5/(3H_0^2),$ $\Omega_\mu=\mu/(H_0^2 a_0^4)$ and $\tau=n H_0t.$

Because of the covariant energy conservation of $\tilde{T}_{\mu\nu}$ (cf. \cite{Binetruy.Deffayet.ea_Brane_cosmological_evolution_2000}) and Bianchi's identity, the energy density can be represented as 
\begin{equation}
\rho=\frac{\rho_{m,0}}{a^3}+\frac{\rho_{r,0}}{a^4} +\frac{\Lambda_4}{\kappa_4^2},
\label{eq:state}
\end{equation} 
where $\rho_{m,0}$ and $\rho_{r,0}$ are constants and $\Lambda_4/\kappa_4^2$ denotes the vacuum energy density. Therefore, the identity 
\begin{equation}
\frac{\rho}{\rho_{\text{crit}}}=\frac{\Omega_m}{x^3}+\frac{\Omega_r}{x^4}+ \Omega_{\Lambda_4} 
\label{eq:rho_rhocrit}
\end{equation}
with the obvious notations for $\Omega_m,$ $\Omega_r$ and $\Omega_{\Lambda_4}$ follows. 

Today, i.e. for $t=t_0,$ it holds that $dx/d\tau=1$ and $x=1.$ Consequently, the identities 
\begin{equation}
\begin{split}
\Omega_{\Lambda_4}&=1-\Omega_k-\Omega_{\Lambda_5}-\Omega_m-\Omega_r \\
& -2\epsilon \sqrt{-\frac{\Omega_{\Lambda_5}}{2}} \sqrt{1-\Omega_k-\frac{\Omega_{\Lambda_5}}{2} -\Omega_\mu},
\end{split}
\label{eq:om4_id}
\end{equation}
and
\begin{equation}
\begin{split}
\Omega_{\Lambda_4}&=-\Omega_m-\Omega_r+\Omega_\mu\\
&+\left( \sqrt{-\frac{\Omega_{\Lambda_5}}{2}} -\epsilon \sqrt{1-\Omega_k-\frac{\Omega_{\Lambda_5}}{2}-\Omega_\mu}  \right)^2 
\end{split}
\label{eq:om4_id2}
\end{equation}
follows from $(\ref{eq:mod_friedmann4})$ and $(\ref{eq:rho_rhocrit}).$
Now we introduce the notations
\begin{equation}
\Omega_h=\Omega_m(x^{-3}-1)+\Omega_r(x^{-4}-1)-\Omega_\mu (x^{-4}-1),
\label{eq:oh}
\end{equation}
and
\begin{equation}
\Omega_\Sigma=\sqrt{1-\Omega_k-\frac{\Omega_{\Lambda_5}}{2} -\Omega_{\mu}}-\epsilon \sqrt{-\frac{\Omega_{\Lambda_5}}{2}}.
\label{eq:omega_sigma}
\end{equation}
Then, the equation $(\ref{eq:mod_friedmann3})$ and an application of the identities $(\ref{eq:rho_rhocrit}),$ $(\ref{eq:om4_id})$ and $(\ref{eq:om4_id2})$ lead to
\begin{equation}
\begin{split}
\frac{\left( \frac{d}{d\tau}x \right)^2}{x^{2}}&=\Omega_k( x^{-2}-1) +\Omega_m (x^{-3}-1) +\Omega_r (x^{-4}-1)\\
&+1 +2\epsilon \sqrt{-\frac{\Omega_{\Lambda_5}}{2}} \left(\sqrt{\Omega_h+\Omega_{\Sigma}^2}-\Omega_{\Sigma}\right).
\end{split}
\label{eq:mod_friedmann5}
\end{equation}
Moreover, it holds that
\begin{equation}
\frac{\left( \frac{d}{d\tau}x \right)^2}{x^{2}}=\frac{\Omega_k }{x^{2}} +\frac{\Omega_{m,\text{eff}} }{x^{3}} +\frac{\Omega_{r,\text{eff}} }{x^{4}}+ \Omega_{\Lambda_{4},\text{eff}},
\label{eq:mod_friedmann6}
\end{equation}
where
\begin{equation*}
\begin{split}
\Omega_{m,\text{eff}}&=\Omega_m \eta,\quad
\Omega_{r,\text{eff}}=\Omega_r \eta- \Omega_{\mu} (\eta-1),\\
\Omega_{\Lambda_4,\text{eff}}&=1-\Omega_k-\Omega_{m,\text{eff}}-\Omega_{r,\text{eff}},
\end{split}
\label{}
\end{equation*}
and
\begin{equation*}
\eta=1+ \epsilon \frac{2\sqrt{-\frac{\Omega_{\Lambda_5}}{2}} \left(\sqrt{\Omega_h+\Omega_{\Sigma}^2}-\Omega_{\Sigma}\right)}{\Omega_h}.
\label{}
\end{equation*}
Notice, that $(\ref{eq:mod_friedmann6})$ is equivalent to $(\ref{eq:mod_friedmann3}),$ but the former equation is available in the original form of the Friedmann equation. The difference arises from the variational coefficients $\Omega_{m,\text{eff}}=\Omega_{m,\text{eff}}(x)$ and $\Omega_{r,\text{eff}}=\Omega_{r,\text{eff}}(x).$ Due to the fact, that the universe, except the earliest period, can be described very well with constant $\Omega_{m,\text{eff}}$ and $\Omega_{r,\text{eff}}$ ($\Lambda$CDM model) the deviations from this description are investigated in the following.
Assuming that $\Omega_\mu$ is bounded from above and that $|\Omega_h|/\Omega_\Sigma^2 \ll 1,$ which is satisfied for sufficiently late times and $\Omega_{\Sigma}^2>0$ or for early epochs if $\Omega_{\Sigma}^2 \gg 0.$ In both cases, the first order approximation of $\eta$ gives
\begin{equation}
\eta \approx \tilde{\eta}=1 +\epsilon \frac{\sqrt{-\frac{\Omega_{\Lambda_5}}{2}}}{\Omega_{\Sigma}},% + \mathcal{O}(\Omega_h).
\label{eq:eta_first_order}
\end{equation} 
which delivers constant approximations for $\Omega_{m,\text{eff}}$ and $\Omega_{r,\text{eff}}$ and converts $(\ref{eq:mod_friedmann6})$ to an effective Friedmann equation. 
In that case, $\Omega_{m,\text{eff}}$ and $\Omega_{r,\text{eff}}$ can be identified with the corresponding values $\Omega_{m,\Lambda CDM}$ and $\Omega_{r,\Lambda CDM}$ of the $\Lambda$CDM model.

Moreover, the term $\Omega_{\mu}(\eta-1)$ in the definition of $\Omega_{r,\text{eff}}$ scales like radiation, modifies the expansion rate during the radiation dominated epoch and causes the nuclear reactions to freeze out at a different temperature. These effects motivates that $|\Omega_{\mu} (\eta-1)| \ll 1.$ 
A more detailed analysis with constraints on light elemental abundances, BBN nuclear reaction rates and the CMB can be found in \cite{Sasankan.Gangopadhyay.ea_Limits_brane-world_2017}.

Then it is of interest to define
\begin{equation}
\Omega_m=\frac{\Omega_{m,\Lambda CDM}}{\tilde{\eta}},\quad \Omega_r=\frac{\Omega_{r,\Lambda CDM}+\Omega_{\mu}(\tilde{\eta}-1)}{\tilde{\eta}},
\label{eq:om_or}
\end{equation}
and
investigate for $\Omega_k=0$ the case
\begin{equation}
\quad\epsilon=-1,\quad  \Omega_{\Lambda_5} \to -\infty,\quad |\Omega_\mu| \ll 1, 
\label{eq:om_limits}
\end{equation}
which generates the desired limits
\begin{equation*}
\Omega_\Sigma \to \infty, \eta \to \frac{1}{2}, |\Omega_{\mu}(\eta-1)| \ll 1. 
\end{equation*}
Using $(\ref{eq:om4_id2})$ and $(\ref{eq:omega_sigma})$ it follows that
\begin{equation}
\Omega_{\Lambda_4}=-\Omega_m-\Omega_r+\Omega_\mu+\Omega_{\Sigma}^2 \to \infty.
\label{eq:limit_ol4}
\end{equation}
From the point of view of quantum field theory, this results is of special interest because it predicts a large vacuum energy density. 
Consequently, the coincidence problem, i.e. $\Omega_m$ and $\Omega_{\Lambda_4}$ are of the same order of magnitude, is avoided and the cosmological constant problem, if not solved, is greatly reduced.

\begin{rem}
The case $|\Omega_h|/\Omega_{\Sigma}^2 \ll 1$ and therefore an effective Friedmann equation can also be realised with $\epsilon=1.$ Using the setting $\delta=2 \Omega_{\mu}/\Omega_{\Lambda_5} >0$ and let 
\begin{equation*}
\delta \gg -\frac{2}{\Omega_{\Lambda_5}} \gg \sqrt{\delta} \gg \Omega_{\Sigma}^2
\end{equation*}  
be fulfilled. Then, it follows that $\Omega_{\Sigma}^2 \gg 1$ and $|\Omega_{\mu}(\eta-1)| \ll 1.$ However, the large value calculations yields $\Omega_{\Lambda_4} \approx -\Omega_m -\Omega_r,$ which doesn't solve the coincidence and cosmological constant problem and will not be considered here.
\end{rem}

In summary, the proposed model of the universe states as follows:
\begin{model}
Taking into account the notations of $(\ref{eq:rho_rhocrit}),$ $(\ref{eq:om4_id2}),$ $(\ref{eq:oh}),$ $(\ref{eq:omega_sigma}),$ $(\ref{eq:eta_first_order}),$ the parameters $(\ref{eq:om_or})$ and $\epsilon=-1.$ Then the evolution of the universe can be described by equation $(\ref{eq:mod_friedmann3}).$ 
\end{model}

In order to complete the model and to evaluate the constraint $(\ref{eq:l_const}),$ the second Friedmann equations which belongs to equations $(\ref{eq:mod_friedmann})$ and $(\ref{eq:cft_einstein})$ are given for $y_{\text{br}}=0$
\begin{equation}
\begin{split}
-\frac{\Lambda_5}{3} +\frac{\ddot a}{a} &=-\frac{1}{12} \kappa_5^4 \tilde{p} \tilde{\rho} 
-\frac{1}{18} \kappa_5^4 \tilde{\rho}^2 - \frac{\Lambda_5}{6} -\frac{\mu}{a^4},\\
-\frac{\Lambda_5}{3}+\frac{\ddot{a}}{a}&=-\frac{\kappa_4^2}{6} \left( \rho+3p \right)-\frac{\kappa_4^2}{12} \left( \tilde{\rho}+3\tilde{p} \right).
\end{split}
\label{eq:second_friedmann}
\end{equation}
Using the equations $(\ref{eq:second_friedmann})$ and after lengthy algebra, one gets a second Friedmann equation without $\tilde{\rho}$ and $\tilde{p}$
\begin{equation}
\frac{\ddot{a}}{a}=-\frac{\kappa_4^2}{6}\left( \rho +3p \right) + \epsilon \frac{\kappa_4^2}{6l}\frac{\rho-3p}{\sqrt{\kappa_4^2 \rho/3-\mu/a ^4}}.
\label{eq:mod_second_friedmann}
\end{equation}
Then, adding $(\ref{eq:mod_friedmann2})$ and $(\ref{eq:mod_second_friedmann})$ and using the notations of this section, it follows that
\begin{equation*}
R=6 H_0^2 \left( \frac{\rho-3p}{2 \rho_\text{crit}}+ \epsilon \sqrt{-\frac{\Omega_{\Lambda_5}}{2}} \frac{\frac{5 \rho-3p}{2\rho_{\text{crit}}}-2 \Omega_{\mu}/x^4}{\sqrt{\frac{\rho}{\rho_{\text{crit}}} -\frac{\Omega_\mu}{x^4}}}\right).
%\label{eq:R}
\end{equation*} 
In section 6 the above results of the RBW model will be analysed with more precise estimates. 

\begin{rem}
The equations $(\ref{eq:mod_friedmann2})$ and $(\ref{eq:mod_second_friedmann})$ contain a sign ambiguity. If $\epsilon=0,$ both equations represents the four-dimen\-sional Einstein equation and can be interpreted as Einstein equation with an additional term in the case of $\epsilon=\pm 1.$ Therefore, there is reasonable hope that an associated four-dimensional action exists, which immediately contains the correct sign and prevent the sign ambiguity. However a derivation of these action is beyond the scope of this paper. 
\end{rem}

\section{Cosmology on the conformal boundary}
\label{sec:Cosm_conf_bound}
As mentioned above, the renormalised Friedmann equation $(\ref{eq:mod_friedmann2})$ is 
equivalent to the $\Lambda$CDM model if $\Omega_{\Lambda_5} \to - \infty.$ 
In this section, this limit is considered from a different perspective. To do so, the gaussian normal coordinates
\begin{equation*}
ds^2=\gamma_{\mu\nu}(x,y)\ dx^\mu dx^\nu +dy^2, \quad \gamma_{\mu\nu}(x,y)=e^{-2y/l}g_{\mu\nu}(x,y),
\label{}
\end{equation*} 
are transformed by $r=l e^{y/l}$ to the coordinates of Fefferman and Graham
$(\ref{eq:coord_Feff_Gra}),$
where the conformal boundary is located at $r=0.$ Hence, the investigated hypersurface $y_{\text{br}}=0$ converges to the conformal boundary of the AdS/CFT correspondence if $l \to 0$ or $\Omega_{\Lambda_5} \to -\infty,$ respectively.
Consequently, the RBW model tends to the $\Lambda$CDM model, which is achieved on the conformal boundary for $l=0.$

The conformal anomaly takes the well-known form
\begin{equation*}
\begin{split}
g^{\mu\nu}T_{\mu\nu}[g]&=\gamma^{\mu\nu}T_{\mu\nu}[\gamma]\\
&=\frac{l^3}{8 \kappa_5^2}\left( R_{\mu\nu}[\gamma]R^{\mu\nu}[\gamma]-\frac{1}{3}R[\gamma]^2 \right)+\mathcal{O}(r^2) \\
&= -\frac{3 l^2 \left( \dot{a}^2+k \right)\ddot{a}}{\kappa_4^2 a^3} +\mathcal{O}(l^\lambda),
\end{split}
\end{equation*}
where the last equation is valid on the brane $y_{br}=0$ (cf. \cite{Bilic_RandallSundrum_versus_holographic_2016}).
Although $r_{\text{br}}=l$ at $y_{br}=0,$ the exponent $\lambda$ is unknown for the moment, since the coefficient of $\mathcal{O}(r^2)$ may also depend on $l.$ 

In order to determine $\lambda,$ equation $(\ref{eq:mod_friedmann2})$ is solved for $\rho.$ Using this identity and equation $(\ref{eq:mod_second_friedmann}),$ one gets 
\begin{equation*}
\begin{split}
T_\mu^\mu&=3p-\rho
= -\frac{3 l^2 \left( \dot{a}^2+k \right)\ddot{a}}{\kappa_4^2 a^3} +\mathcal{O}(l^2), \quad \epsilon=1,\\
T_\mu^\mu&=3p-\rho
= -\frac{48}{\kappa_4^2} l^{-2} +\mathcal{O}(l^0),\quad \epsilon=-1,
\end{split}
\end{equation*}
The last result can also be obtained by using equation $(\ref{eq:state})$ and $\Lambda_5 \approx -\Lambda_4/2$ which follows from equation $(\ref{eq:om4_id2})$ for sufficient small $l$ and $\epsilon=-1.$  
Therefore, the equation of state $(\ref{eq:state})$ is compatible with the conformal anomaly. Since the next section estimates the parameter $l$ by cosmological observations, the last equation is suitable for an estimation of the conformal anomaly.

%Hence, the conformal anomaly of the renormalised brane world model approaches infinity if the brane tends to the conformal boundary.

\section{Describing the cosmological history}

In order to assess the RBW model, the ability to describe cosmological observations is examined in this section. Therefore, the deviations between the models RBW and $\Lambda$CDM are considered.
First, in the view of $|\Omega_h|/\Omega_\Sigma^2 \ll 1,$ we state that the higher the redshifts $z=x^{-1}-1,$ the larger is the difference between the RBW and the $\Lambda$CDM model or, in other words, the larger $\Omega_\Sigma,$ the further the agreement between the RBW model and the $\Lambda$CDM model goes back in time.
%Therefore, the RBW model and a sufficiently large $\Omega_\Sigma$ can describe the cosmological observations at least as well as the $\Lambda$CDM model.

According to $(\ref{eq:limit_ol4})$ and using the constraint $|\Omega_\mu(\eta-1)| \ll1$ one can conclude, the larger $\Omega_{\Lambda_4}$ and therefore the vacuum energy density, the better is the agreement between the RBW model and the $\Lambda$CDM model!

It follows from the above arguments, that it is adequate to measure the differences between both models only at high redshifts.
In order to do that, $z \gg 1$ gives the approximation 
\begin{equation*}
\frac{H(z)^2}{H_{\Lambda CDM}(z)^2} \approx \frac{\eta}{\tilde{\eta}}-\frac{\Omega_\mu}{\Omega_{r,\Lambda CDM}} \left( \frac{\eta}{\tilde{\eta}}-1 \right).
\label{}
\end{equation*} 
%where $H(z)=\dot{x}/x$ is the variational Hubble parameter and $H_{\Lambda CDM}(z)$ denotes the analogous parameter from the $\Lambda$CDM model.
Using the second order approximation 
%w.r.t. $|\Omega_h|/\Omega_\Sigma^2$ 
of $\eta$ and $1-\Omega_\mu \ll -\Omega_{\Lambda_5}/2,$ which is motivated by $(\ref{eq:om_limits}),$ it follows that
\begin{equation*}
%\frac{H(z)^2}{H_{\Lambda CDM}(z)^2} \approx 1+(1-\Omega_\mu) \frac{\sqrt{-\frac{\Omega_{\Lambda_5}}{2}}}{4 \sqrt{1-\frac{\Omega_{\Lambda_5}}{2}-\Omega_\mu}} \frac{\Omega_h}{\Omega_{\Sigma}^2},
\frac{H(z)^2}{H_{\Lambda CDM}(z)^2} \approx 1+\left(1-\frac{\Omega_\mu}{\Omega_{r,\Lambda CDM}}\right) \frac{\Omega_h}{4\Omega_{\Sigma}^2}.
\label{}
\end{equation*} 
Conversely, by specifying the maxiumum deviation between both models, the corresponding redshift range $[0,z^{*}]$ can be determined in which both models represent all cosmological observations equally well. Namely, let $\zeta=\Omega_h/\Omega_{\Sigma}^2$ be a positive and fixed value and using equation $(\ref{eq:oh}),$ $\zeta$ is connected with the upper limit 
\begin{equation*}
z^{*}(l) \approx\left( \frac{2\zeta |\Omega_{\Lambda_5}|}{\Omega_r-\Omega_\mu} \right)^{1/4}.
\end{equation*}
Notice that the constraint $\Omega_r > \Omega_\mu$ is compatible with $|\Omega_\mu (\eta-1)|\ll 1.$
%If $\zeta$ is small, then it follows that $H(z^{*})^2/H_{\Lambda CDM}(z^{*})^2\approx 1+(1-\Omega_\mu)\zeta/4.$
For example, if $\Omega_\mu=0$ and the big bang nucleosynthesis (BBN) at $z_{\text{BBN}}=6 \times 10^9$ should be represented with the RBW model, then 
\begin{equation*}
6 \times 10^9 \le z^{*}(l)
%\label{eq:l_const2}
\end{equation*}
is the second constraint for $l$ next to $(\ref{eq:l_const})$ with $r_{\text{br}}=l.$
Choosing $z^{*}=6 \times 10^9$ as a first guess and the desired deviation, e.g. $H(z)^2/H_{\Lambda CDM}(z)^2\approx 1.01,$ implies with $\Omega_{r,\Lambda CDM}=0.00005$ that $\Omega_{\Lambda_5}=-1.62 \times 10^{36}.$
Using also $H_0=67.66\, \text{km} \text{s}^{-1} \text{Mpc}^{-1},$ $\Omega_{m,\Lambda CDM}=0.3111$ and $\Omega_k=0,$ the AdS-parameter $l=1.11 \times 10^{-18}/H_0=1.52 \times 10^8 \text{m}$ turns out to be too large due to $|R(z_{\text{BBN}},l) l^2 \log(l^2)|=3.13 \times 10^{16}.$

Next, selecting $z^{*}=2 \times 10^{13}$ the following values are obtained in an analogue manner: 
$\Omega_{\Lambda_5}=-2 \times 10^{50},$
$l=10^{-25}/H_0=13.67 \text{m},$
$|R(z_{\text{BBN}},l) l^2 \log(l^2)|=0.0009$ which is valid w.r.t. constraint $(\ref{eq:l_const}).$
\begin{figure}[!htbp]
\begin{center}
\includegraphics[width=1.0\columnwidth]{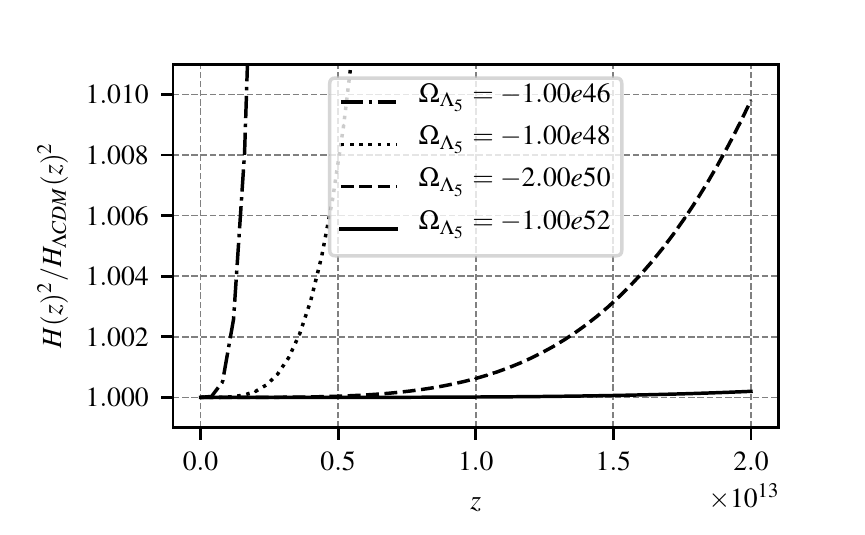}
\caption{The deviations between the RBW and the $\Lambda$CDM models for exemplary selected $\Omega_{\Lambda_5}.$}
\label{fig:deviations}
\end{center}
\end{figure}
Figure \ref{fig:deviations} validates the above analysis and shows the characteristics of the deviations between the RBW and the $\Lambda$CDM model for redshifts $z<2 \times 10^{13}.$ 

In summary, the parameter selection $l \precsim 10 \text{m}$ guaranties that the RBW model requirements and the agreements with observations are given up to the big bang nucleosynthesis.

Moreover, if the vacuum energy density is defined by quantum field theory, that is $\Omega_{\Lambda_4} \approx 10^{122},$ then, because of $\Omega_k=0$ and $1-\Omega_\mu\ll -\Omega_{\Lambda_5}/2,$ we obtain $\Omega_{\Lambda_5} \approx -\Omega_{\Lambda_4}/2$ by using $(\ref{eq:om4_id2}).$ A similar procedure as before provides that $l$ is approximately equal to the Planck length $l_P,$ which is far below the above upper bound of $l \precsim 10 \text{m}$ and therefore guaranties the agreement with observations.

According to the AdS/CFT description, $\Omega_{\Lambda_5} \approx -10^{122}$ restricts the fifth dimension of the bulk space-time to 
%\begin{equation*}
%r_{\text{br}}=\frac{\sqrt{-2/\Omega_{\Lambda_5}}}{H_0} \leq r <\infty
%\end{equation*}
$l_P \leq r <\infty.$

\section{Concluding remarks}
In this paper, it has been demonstrated that the RBW model, which equals the $\Lambda$CDM model on the conformal boundary for $l=0,$ is more evident than the original $\Lambda$CDM model. 
This conclusion has been confirmed by the facts that the observational quality of the $\Lambda$CDM model is preserved, no fine tuning of the brane is necessary, the coincidence problem is avoided and the cosmological constant problem, if not solved, is greatly reduced.

Hence, there is a reasonable hope that if the RBW model is confronted with the cosmic microwave background data from the Planck measurements the tension between the early and late universe can be reduced and that the Planck data can determine the vacuum energy density from quantum field theory. 

Moreover, the above conclusions change the complete picture of our universe and enables further applications of the AdS/CFT correspondence in cosmology.

%%\section*{References}

%\bibliographystyle{unsrt}
%\bibliography{EEbib4}

\end{document}